# Size-Dependent Lattice Pseudosymmetry for Frustrated Decahedral Nanoparticles

Oliver Lin[1], Zhiheng Lyu[2], Hsu-Chih Ni[2], Xiaokang Wang[2], Yetong Jia[3], Chu-Yun Hwang[2], Lehan Yao[2], Jian-Min Zuo[2,4], Qian Chen[1,2,4,5,6]∗

[1] Department of Chemistry, University of Illinois, Urbana, IL 61801, U.S.A.
[2] Department of Materials Science and Engineering, Grainger College of Engineering, University of Illinois, Urbana, IL 61801, U.S.A.
[3] School of Mechanical Engineering, Purdue University, West Lafayette, IN 47907, U.S.A.
[4] Materials Research Laboratory, University of Illinois, Urbana, IL 61801, U.S.A.
[5] Department of Chemical and Biomolecular Engineering, University of Illinois, Urbana, IL 61801, U.S.A.
[6] Beckman Institute for Advanced Science and Technology, University of Illinois, Urbana, IL 61801, U.S.A.
*Correspondence should be addressed to: qchen20@illinois.edu

**Abstract**

Geometric frustration is a widespread phenomenon in physics, materials science, and biology, occurring when a system's geometry prevents local interactions from being all accommodated. The resulting manifold of nearly degenerate configurations can lead to complex collective behaviors and emergent pseudosymmetry in diverse systems such as frustrated magnets, mechanical metamaterials, and protein assemblies. In synthetic multi-twinned nanomaterials, similar pseudosymmetric features have also been observed and manifest as intrinsic lattice strain. Despite extensive interest in the stability of these nanostructures, a fundamental understanding remains limited due to the lack of detailed structural characterization across varying sizes and geometries. In this work, we apply four-dimensional scanning transmission electron microscopy (4D-STEM) strain mapping over a total of 23 decahedral NPs with edge lengths ($d$) between 20 and 55 nm. From maps of full 2D strain tensor at nanometer spatial resolution, we reveal the prevalence of heterogeneity in different modes of lattice distortions, which homogenizes and restores symmetry with increasing size. Knowing the particle's crystallography, we reveal distinctive spatial patterns of local lattice phase transformation between face-centered cubic (*fcc*) and body-centered tetragonal (*bct*) symmetries, with a contrast between particles below and above $d$ = 35 nm. The results suggest a cross-over size of the internal structure occurs, as particles shape transition from modified-Wulff shape favored at nanoscale to faceted, pentagonal bipyramidal shape. Ultimately, our 4D-STEM mapping provides new insight to long-standing mysteries of this historic system and can be widely applicable to study nanocrystalline solids and material phase transformation that are important in catalysis, metallurgy, electronic devices, and energy storage materials.

**Introduction**

Geometric frustration occurs widely in physics, materials science, and biology when the system's geometry precludes the simultaneous satisfaction of all local interactions, producing complex structures and exotic properties. Examples can include all the way from emergence of chirality for polarization sensitive imaging, the spin ice states in ferromagnetic Kagome lattices for quantum computing to the origami-inspired surface with adjustable compressibility as robotics[1,2]. In crystalline systems, accommodating geometric frustration can induce slight distortion from the ideal lattices of long-range order and lead to *pseudosymmetry*, a phenomenon recognized decades ago as an evolutionary adaptation in biological structures[3,4]. For instance, neuraminidase of the influenza virus exhibits sliding domains that generate pseudosymmetry. The resulted structural flexibility is capable of surpassing the limitations of the lock-and-key mechanism and to more effectively evade host's immune system[5].

Pseudosymmetry also exists in synthetic nanomaterials though it has been much less discussed[6]. On one hand, geometrically frustrated crystals such as five-twinned metallic seed particles serve as the cornerstone for the colloidal synthesis of a wide range of shape-controlled nanoparticles (NPs), such as rods, wires, and bipyramids of gold, silver, and copper[7–9]. These NPs exhibit remarkable catalytic, plasmonic, mechanical, and chiroptical properties, great for applications in optoelectronics, bioimaging and biosensing, fuel cells, and machine vision[10–13]. Pseudosymmetry is expected to emerge in five-twinned NPs because most metals have face-centered cubic (*fcc*) lattice structure; imposing the five-fold rotational symmetry onto *fcc* lattice will cause atomic displacements in the NPs, altering the local lattice symmetry, strain, and functions of NPs. As a prominent example, unrelaxed strain at the twin boundaries of NP catalysts can modulate the energy of *d*-band electrons and substantially enhance the catalytic activity and selectivity[14]. On the other hand, the understanding of the structural distortion and pseudosymmetry at the nanoscale remains scant. For the five-twinned NPs[15,16], early studies suggested that the geometric frustration is stabilized by forming low energy {111} facets to minimize surface energy and accommodate the energetic cost of bulk strain and twinning.[17] Continuum models further showed that surface energy and volumetric strain energy scale to $r^2$ and $r^3$ respectively, where $r$ is the particle radius assuming a spherical shape. This scaling law difference suggests a particle size-dependent pseudosymmetry behavior, which remains unexplored. More detailed modeling of the internal structural distortion of NPs and its impact on properties is challenging[18] because colloidally-synthesized five-twinned NPs are complex, with variations in particle size (from a few nanometers (nm) to hundreds of nm), particle shape (from perfect decahedra, truncated Marks decahedra, pentagonal rods, to those with asymmetric domains)[19,20], metal composition, and ligand adsorption energies on different facets. Previous experimental studies focused on the NPs of edge length (*d*) smaller than 10 nm, a size range that allows imaging of atoms or lattice fringes by scanning/transmission electron microscopy (S/TEM). Yet many geometrically frustrated NPs are larger and are currently precluded in these studies.[21,22] There remain extensive interest and a lack of fundamental understanding regarding how these particles sustain and stabilize the structure against strain energy and twinning across a wide NP size range.

Here we present the first study of *particle-size dependent* crossover of pseudosymmetry behaviors in multi-twinned NPs, using a total of 23 five-twinned gold decahedral (Dh) NPs as a focus system (Method), which are large, with edge length *d* ranging from 20 to 55 nm, and hard to see through quantitatively using the imaging mode of S/TEM. We develop a workflow integrating four-dimensional scanning transmission electron microscopy (4D-STEM) capable of imaging thick samples, diffraction pattern (DP) analysis following the principles of solid mechanics and crystallography, to map the local strain, lattice displacements and symmetry of NPs at the otherwise inaccessible nm resolution. Gold Dh NP shapes have a $D_{5h}$ point group and accommodate five tetrahedral "grains" around a shared axis, resulting in an angular geometric misfit known as a disclination of 7.35°. By analyzing the DPs from 4D-STEM (Figure 1a), we map strain tensor ($\varepsilon_{220}$, $\varepsilon_{002}$ and shear strain $\gamma$) and rigid body rotation ($R$) of local lattices of each grain while considering the full complexity of the five-fold twin. Solid mechanics principles allow us to merge the five grains to a global coordinate and relate strain map to the spatial distribution of lattice displacement gradient which we are able to map directly from 4D-STEM. These comprehensive whole-particle maps show that all our NPs, regardless of size, are strained heterogeneously at the nm scale, maintaining normal strains as averaged at 2% without the formation of extended defects like dislocations or nanotwins. The NPs exhibit size-*in*dependent spatial patterns of $\gamma$ and $R$ to close the 7.35° geometric gap. Meanwhile, size-dependent behaviors also emerge. Large Dh NPs ($d > 35$ nm) exhibit faceted pentagonal edges and what we refer to as the "bulk" behavior, where $\varepsilon_{220}$ and $\varepsilon_{002}$ show clear *intra-grain* heterogeneity, with strain focused at particle edges, and minor *inter-grain* heterogeneity. Both features are consistent with a finite element analysis (FEA) modeling of a bulk perfect decahedron. In contrast, $\varepsilon_{220}$ and $\varepsilon_{002}$ of small Dh NPs ($d < 35$ nm) are uniform in each grain, but have large inter-grain heterogeneity, revealing a "nanosize" effect that we attribute to early-stage grain merging pathways during NP synthesis that is not covered in previous modeling of strain field[23]. This cross-over from bulk to nano behaviors with decreasing particle size correlates to pseudosymmetry, which we define and measure from the crystal structure of Au. While the NP lattice symmetry is predominantly body-centered tetragonal (*bct*) lattice expected for five-twinned NPs[24], we observe an intriguing emergence and spatial redistribution of regions of bulk *fcc* symmetry as the particle size increases, to restore the ideal symmetry under strain.

Our work on five-twinned NPs of a wide size range elucidates how the internal pseudosymmetric distortion embeds in nanoscale heterogeneity and exhibits a crossover to bulk behaviors. As geometric frustration has been found beyond *fcc* noble metals, such as intermetallics of $Au_3Cu$ and FePt in face-centered tetragonal structure with disclination, our understanding will be related to these other phase and strain engineered NP-based functional materials[25–27]. Our 4D-STEM strain and symmetry mapping also extend to broadly the understanding of twin boundary formation, strain relaxation, and phase transformation in other materials used as catalysis, mechanical absorptive materials, electronic devices, and energy storage materials.

**Results**

**4D-STEM mapping of strain tensor components in gold Dh NPs**

Gold Dh NPs are colloidally synthesized using seed-mediated growth[9], where the NP sizes are controlled by the seed amount (Method, Figure 1b). All the NPs maintain five-fold twins over the size range of $d = 20$ to 55 nm. Closing the geometric gap requires lattice straining and structural distortion. As such, we expect to observe predominately *bct* lattice, along with the redistribution of local lattice symmetry of *fcc* throughout the NPs (Figure 1b). We find ideal {111} facets in the NPs without further faceting, unlike the Marks decahedron more commonly observed in large Pd NPs.[28] The NPs transition from decahedra of highly rounded edges to sharp-edged pentagonal bipyramidal shape as their sizes increase.

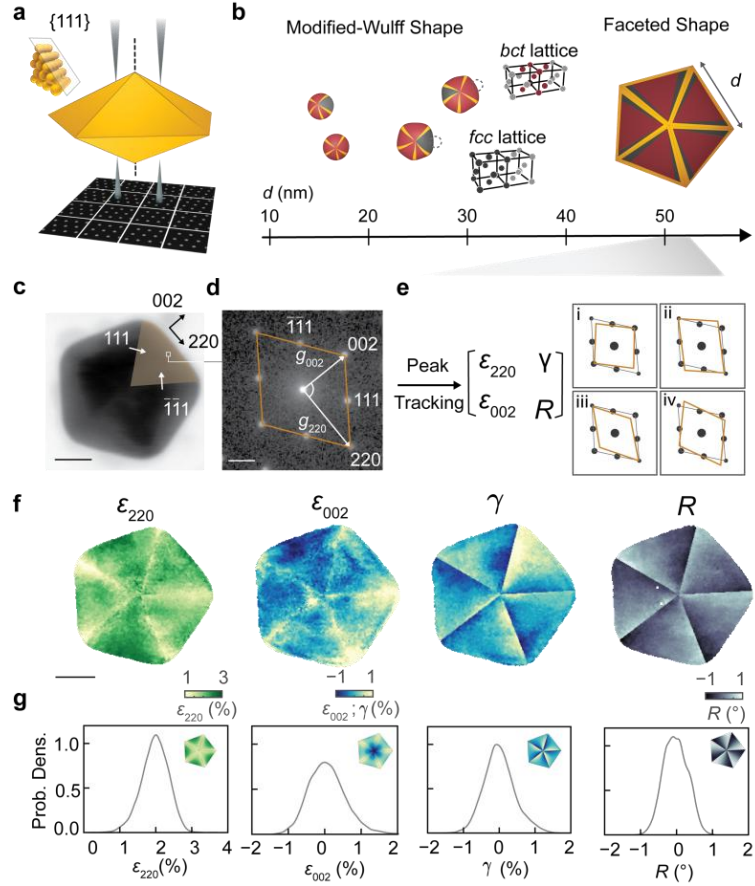

**Fig. 1 Mapping of heterogeneous strain distributions in colloidally synthesized Au Dh NPs by 4D-STEM.** (**a**) Schematics of 4D-STEM measurement configuration that aligns incident electron beam to the particle disclination axis of a gold Dh NP. (**b**) Illustration of how the particle shape and symmetry of Au Dh NPs change as the edge length $d$ increases from 20 to 55 nm. (**c**) VBF of a particle of $d = 50.1$ nm, with a grain highlighted to show its local coordinate. (**d**) DP of the pixel boxed in (**c**), labeled with index of each diffraction discs, as well as the $g$-vectors and angle in between. (**e**) The types of lattice distortion modes can be identified by tracking the positions of diffraction discs, (i) normal strain in (220) direction ($\varepsilon_{220}$), (ii) shear strain ($\gamma$), (iii) normal strain in (002) direction ($\varepsilon_{002}$), and (iv) lattice rotation ($R$). (**f**) Experimental strain maps of the same Au Dh NP in (**c**). (**g**) Histograms of strain values from experimental maps. Insets: FEA simulated strain fields. Scale bars, images (c,f): 20 nm, DP (d): 5 nm$^{-1}$.

To map the components of strain tensors, 4D-STEM is used to collect DPs at a 1.7 nm electron probe size, 0.376 nm step size, and a convergence angle of 0.6 mrad. The partial overlapping of sampling regions between neighboring steps is intentional to ensure continuity for displacement field mapping. We align the particle to the disclination axis, so all the tetrahedral grains are along the [110] axis to map the complete strain tensor components. Figure 1c shows a typical virtual bright field (VBF) image of a large particle ($d = 50.1$ nm). As the five grains are positioned in five-fold rotational symmetry, the crystallographic direction in each grain is rotated relatively in-plane. The (220) and (002) lattice plane directions—parallel and perpendicular to the grain edge respectively—are the orthogonal basis that defines the local coordinate for strain mapping. Using the DP (Figure 1d) at one pixel of VBF (labeled in Figure 1c) as an example, strain mapping tracks the Bragg peak positions of available diffraction discs by circular Hough transform method[29], fits into an experimental lattice, and then compares with a global reference lattice of bulk gold (Method and SI). The normal strain ($\varepsilon_{220}$ and $\varepsilon_{002}$) based on the lengths of *g*-vectors ($g_{220}$ and $g_{002}$), the shear strain ($\gamma$, change of the angle between the *g*-vectors relative to the angle for an ideal *fcc* lattice), and the in-plane rotation (*R*) are extracted from the DPs at each pixel, corresponding to different distortion modes in reference to an ideal lattice (Figure 1e). Because each grain has a different local coordinate, the strain analysis is performed five times, masked, and combined to generate the strain maps of the entire particle with four components (Figure 1f).

The strain maps (Figure 1f) reveal nanoscopic heterogeneity for all the strain tensor components. In the example of the large particle (*d* = 50.1 nm), the $\varepsilon_{220}$ map shows that the (220) lattice plane direction is all tensile strained, more at the particle edge (averaged at 2%) than at the twin boundary. In comparison, for the (002) lattice plane direction, compressive strain occurs at particle edges (–1%) with tensile strain at twin boundary (1%). Both $\gamma$ and *R* maps show alternating patterns within each grain over the range of –1% to 1%, with the extrema occurring near twin boundaries. Figure 1g lists the histograms of the four strain components, the overall range matching with a figure-of-merit expectation of a 7.35° geometric gap, about 2% of the entire 360° angular space. FEA simulations of the same set of strain tensor components of a perfectly shaped Au decahedron (insets in Figure 1g) have patterns matching qualitatively with our experiments, although the latter show more heterogeneity, suggesting that the projected volumetric strain field of the large particle can be modeled using bulk anisotropic elasticity to close the 7.35° gap. FEA models the 3D particle shape with the strain tensor components projected and averaged along the same direction as in our experiment for comparison (Method).

The 4D-STEM-based strain mapping provides comprehensive measurement of local lattice distortion at the nm spatial resolution, for samples of thickness up to ~300 nm. Our convergence of all five grains using one global reference reveals the nanoscale heterogeneity and the long-range effect of lattice straining across grains. In comparison, previous studies of small Dh NPs (*d* <10 nm) used aberration-corrected TEM imaging to apply geometric phase analysis (GPA) to high-quality electron microscopy images, obtaining $\gamma$ and *R* of the same alternating patterns as ours to explain gap closing but not the other components[30]. These patterns were attributed to the nature of the structural distortion based on disclination. We show that the same mechanism applies to large NPs. In the more recent study using atom tomography for Dh NPs of *d* ~ 5 nm, the 3D atomic coordinates of NPs were directly used to calculate $\varepsilon_{220}$ in one grain, showing tensile strain at the edge[31], consistent with Figure 1f, but not for the complete mapping of all the five grains to see intra- or inter-grain heterogeneity.

**Visualization of geometric gap closing**

To visualize lattice distortions in real space that close the geometric gap, Figure 2a summarizes symbolic presentations of distortion modes. Starting from a reference *fcc* lattice (undistorted bulk gold) viewed from [100] direction, which we denote as a perfect rectangle, at the center of the particle edge, there are minimal $\gamma$ and *R* lattice shear (denoted by ○), but large $\varepsilon_{220}$ as tensile strain and $\varepsilon_{002}$ as compressive strain (distorted rectangles). Near twin boundaries, $\varepsilon_{220}$ is reduced and $\varepsilon_{002}$ turns tensile, with severe angular distortion visible in $\gamma$ and *R*, denoted as a sheared rhomboid. At the two twin boundaries of one grain, the directions of the angular distortion are opposite, as if the grain is bent. High-magnification high-angle annular dark-field (HAADF) STEM images (Figure 2b, more in SI) show that the five grains separate the total 360° unevenly, suggesting the importance of studying the NP as a whole with all the five grains resolved and connected.

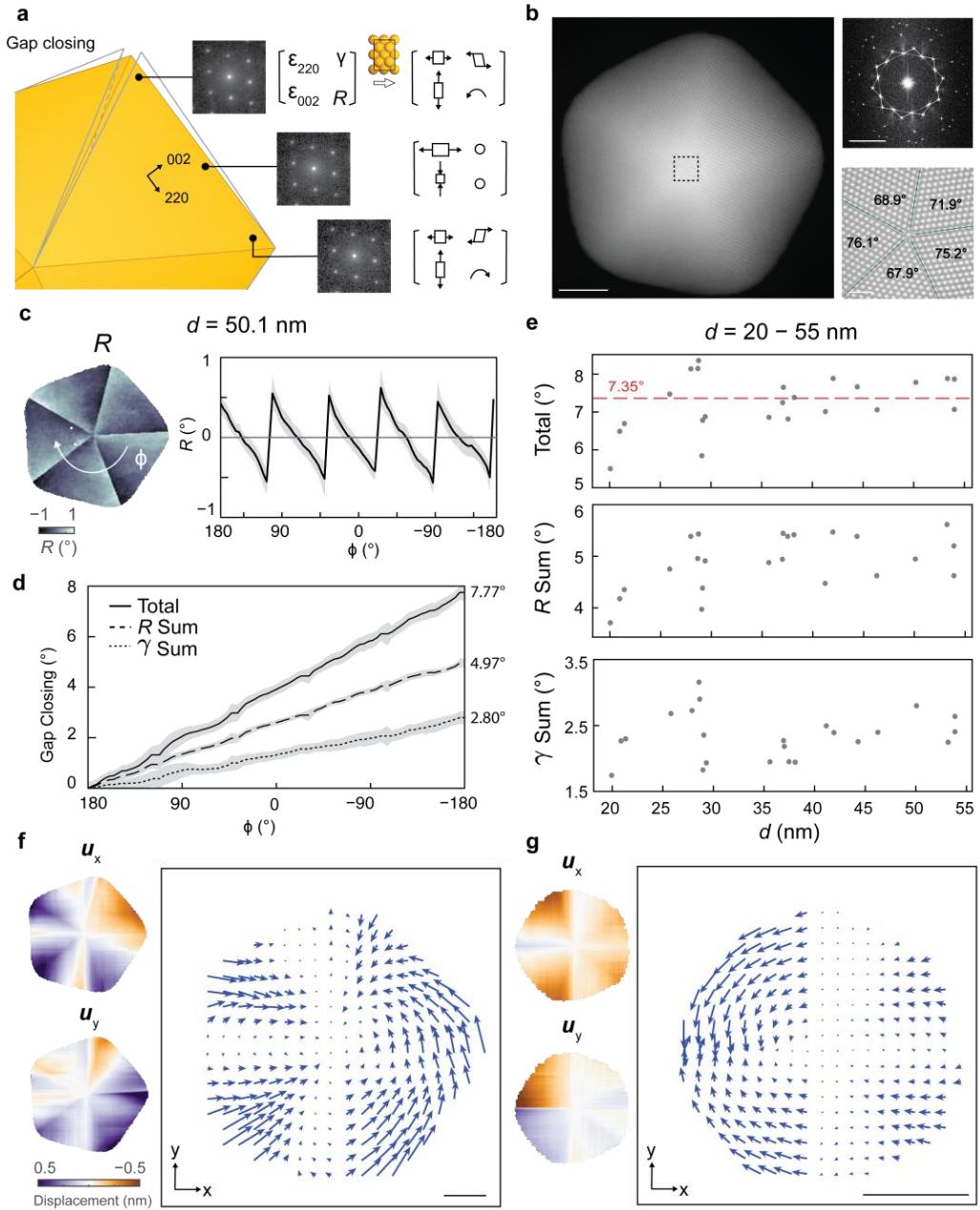

**Fig. 2 Geometric gap closing and heterogeneity.** (**a**) Schematic of spatially varying lattice distortion in different modes to fulfill the intrinsic geometric gap imposed by five-fold twin structure using the DPs of different parts of the NP shown in Figure 1c−g ($d$ = 50.1 nm) as an example. (**b**) Ac-corrected STEM image of a particle at $d \sim 35$ nm with its fast Fourier transform on the top right showing the lack of large-scale defects. The zoom-in at the disclination axis showing uneven distribution of angles by five grains. (**c**) The radial averaging of the alternating patterns of $R$ at $d$ = 50.1 nm is plotted over the azimuthal angle ($\phi$) and summed to compute the values in (d) and the shaded areas are the standard deviation. (**d**) The sum of R and γ to quantify the angular distortion towards the geometry gap. (**e**) Scattered plot all the summed values over the particle studied at different sizes. (**f**) (**g**) The displacement vector field maps in global Cartesian coordinate system with the magnitude maps on the left. Scale bars: 10 nm. (f) large particle, $d$ = 50.1 nm; (g) small particle, $d$ = 20.5 nm.

The accumulated lattice distortion in terms of angles to close the gap is calculated from the $\gamma$ and $R$ maps. Using the particle in Figure 1c–g as an example, because the lattice rotation features mostly vary not radially but azimuthally, the values of $R$ can be averaged along the radial direction to collapse the two-dimensional (2D) map (Figure 1f) into a plot (Figure 2c) along an angle ($\phi$) from –180° to 180°. $R$ values exhibit alternating pattern, smoothly varying within a grain and have sharp transitions across twin boundaries. The maximum $R$ varies across the grains. To calculate the lattice distortion for the entire particle, the same plotting is applied to $\gamma$ and the amount of distortion inside each grain is summed (Figure 2d). This workflow has been used to the 23 differently sized NPs (Figure 2e). Such survey and sampling of a series of NPs show a plateau around the theoretically expected value of 7.35° when $d > 35$ nm. When $d < 35$ nm, the total sums fluctuate around 7.35° mostly by ~1°, suggesting that gap closing through lattice rotation still works, although the radially averaged values might not be the most accurate due to more complex heterogeneity in small NPs (SI).

While $\gamma$ and $R$ maps quantitatively capture the lattice rotation of each grain using the local coordinate, the 2D displacement field maps offer a holistic visualization of the lattice displacements for the entire particle, for the first time using DPs. In our 4D-STEM experiment, the pixel size in a strain map is known and small (0.376 nm). Thus, the strain tensor maps can be related to displacement field maps, showing how each pixel (not an atom) of a continuous solid moves to close the geometric gap (SI). Because the NPs are oriented to have mirror symmetry on x-y plane during DP collection, we see such 2D displacement is representative for the NP. By setting the center of Dh NP as zero displacement, we plot the continuous displacement field maps as a vector with two components of $\boldsymbol{u}_x$ and $\boldsymbol{u}_y$ along the global x and y axes (Figure 2f,g) for two NPs of different sizes, showing distinctive features. For the small NP ($d = 20.5$ nm; Figure 2g), one single geometric gap closure is shown at the twin boundary at the left side of the particle, where the grains above and below this twin boundary have movements towards it, showing intra-grain heterogeneity. In comparison, the large NP's displacement field ($d = 50.1$ nm; Figure 2f) shows closure of multiple distributed gaps. The common observation that the closure of gaps is not uniformly distributed among all grains can be originated from the various pathways of forming five-fold twins during colloidal synthesis, such as successive twining and layer-by-layer growth[32], leading to grain-to-grain variation.

Such visualization of displacement fields is made possible by compiling the displacement vectors from the particle center over all the five grains with the same reference and coordinate. The maps consider the full strain tensor at a global coordinate system, not by tracking 3D single atoms which are inaccessible with large NPs, in spirit aligned with the particle image velocimetry analysis used to map collective motions of particles without localizing each of them in optical microscopy.

**NP size-dependent strain distribution and pseudosymmetry: intra- and inter-grain heterogeneity**

We extend the strain analysis above to the Dh NPs of different sizes, where the spatial distributions of lattice distortion exhibit a nano to bulk crossover. Figure 3a shows the HAADF images and the strain maps of Dh NPs with $d$ varying from 20 and 55 nm. The alternating patterns in $\gamma$ and $R$ maps across five grains sustain for all the NPs, confirming the size-*in*dependence of the gap closing mechanism as discussed earlier. Yet, the maps of normal strain components ($\varepsilon_{220}$ and $\varepsilon_{002}$) show a clear NP size dependence. As noted in Figure 2e, $d = 35$ nm separates mildly the regions of consistent total rotation angles. Here, as a nice coincidence, it also defines the size regions for normal strain components. For small particles ($d < 35$ nm), the $\varepsilon_{220}$ and $\varepsilon_{002}$ distribution in each grain shows more of a diffusive feature, meaning one grain is homogeneously strained, yet the values of strain can be different among grains up to 1%. In other words, the particles have significant *inter*-grain heterogeneity while minor *intra*-grain heterogeneity. In contrast, the large particles ($d > 35$ nm), one additional example being the NP shown in Figure 1, have comparatively minor *inter*-grain heterogeneity and significant *intra*-grain heterogeneity. Quantitatively, Figure 3b (i & iii) shows the profiles of $\varepsilon_{220}$ of a selected grain along the path of particle center-to-edge: small particle exhibits steady increase in $\varepsilon_{220}$ while that for large particle fluctuates and then increases towards the particle edge. Figure 3b (ii & iv) shows the profiles of $\varepsilon_{220}$ along the particle boundaries, revealing significant dip near twin boundaries for the large particle although the profiles for each grain are similar, while the small particle shows a step-like pattern signaturing significant different values for two grains (from −1% to 2%).

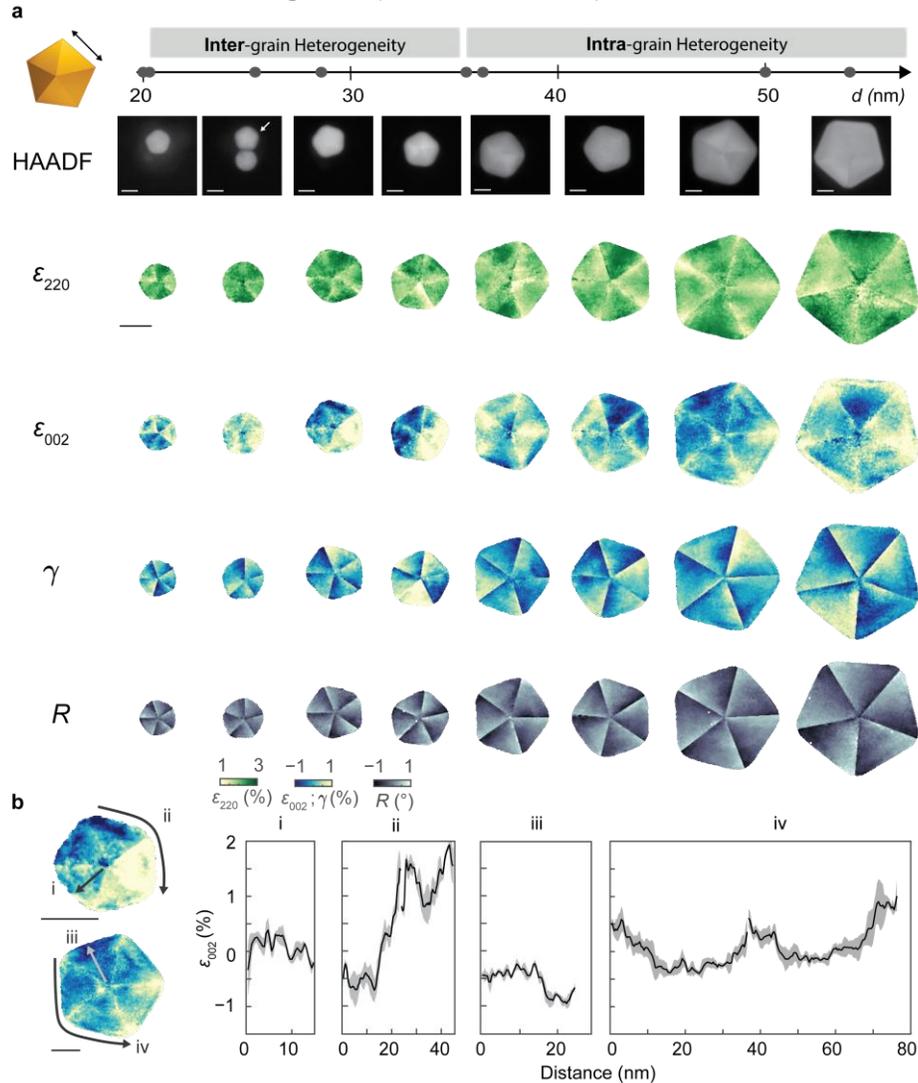

**Fig. 3 Particle size dependence of strain distribution for each component and lattice rotation. (a)** Particle morphologies shown by μp-STEM HAADF images and strain maps of each strain component of particles at $d$ between 20 to 55 nm **(b)** Strain profiles of small (i, ii) and large particles (iii, iv) in center-to-edge (i, iii) and tip-to-tip (ii, iv) paths. Scale bars = 20 nm.

Note that all the NPs on the ensemble level exhibit similar ranges of strain values; it is the spatial patterns of the strain values that show size-to-size, grain-to-grain, pixel-to-pixel variations, which can only be captured quantitatively and comprehensively by our workflow. To understand and quantify such local lattice symmetry variation, we define and compute the *k* parameter, namely the *pseudosymmetry* parameter, the ratio of *d*-spacing along two lattice plane directions ($d_{220}$ and $d_{002}$)[18], based on the normal strain components and particle crystallography,

$$k = \frac{d_{220}}{d_{002}} = \frac{1}{\sqrt{2}}\left(\frac{1+\varepsilon_{220}}{1+\varepsilon_{002}}\right) = \tan\theta \quad (1)$$

where 2θ is the interplanar angle between two (111) twinned directions when the Au lattice is at [110] zone axis (Figure 4a). In an ideal *fcc* lattice, 2θ is exactly 70.53°. For a five-twinned NP, it has been postulated that if all the five grains evenly divide the space of 360°, 2θ must become 72° to close the gap, and this angular distortion will break the symmetry from *fcc* through a tetragonal transformation into *bct* lattice (Figure 4a)[16]. This transformation can be measured by the *k* parameter (defined in Eqn. 1) as a continuous description of the extent of symmetry breaking, ranging from no symmetry breaking (0.7071 for the bulk *fcc* symmetry) to evenly dividing the space (0.7265 for *bct*) following Eq. 1 across the entire particle at nm spatial resolution.

As such, Figure 4b summarizes the *k* parameter maps of 16 particles with varied sizes. The trend starts with the observation of homogeneous distribution of *bct*-like symmetry across the entire particle for small NP size (row 1, Figure 4b; *k* map predominantly red). Then, *fcc*-like symmetry emerges, mostly within one grain, at the twin boundaries and edges (row 2, Figure 4b). Subsequently, the particle level *k* patterns appear more severely distorted (as darker red and black) while *fcc*-symmetry start spreading to more than one grain (row 3, Figure 4b), corresponding to *d* ~ 35 nm crossover of heterogeneity discussed in Figure 3. Lastly, *fcc*-like symmetry appears almost evenly near the five twin boundaries and edges while the *bct*-symmetry regions stay within grains (row 4, Figure 4b), where inter-grain heterogeneity reduces with less distorted regions. When we binarize the *k* histogram at exactly the mid-point, the normalized population of *fcc*-like phase never exceeds 40% (Figure 4c) before the distribution starts to homogenize. For particles smaller than our size range, we could extrapolate that they ought to have more homogeneous distribution of *bct* symmetry across the entire particle.

For the first time, we reveal the lattice pseudosymmetry of geometrically frustrated NPs is spatially varying, and that its distribution is size-dependent. Though projection-based 4D-STEM cannot tell tetragonal symmetry apart from orthorhombic symmetry that has been postulated in literature[16], the assignment of *bct* lattice in I4/mmm space group is consistent with recent literature of using X-ray diffraction to assign symmetry of highly monodispersed Au Dh NPs at similar sizes (31 nm and 49 nm therein)[24]. Wu *et al.* applied the analysis of atomic HAADF for very small (*d* < 5 nm) particles[33]. When referring to the strain maps, it can be observed that most of the *fcc*-regions have tensile strain in both normal directions, meaning the volume of the local lattice is in fact larger. Furthermore, because γ is on the same order of magnitude as ε components, the symmetry is in fact distorted, to base-centered monoclinic symmetry, further highlighting the pseudosymmetric nature of Dh particles.

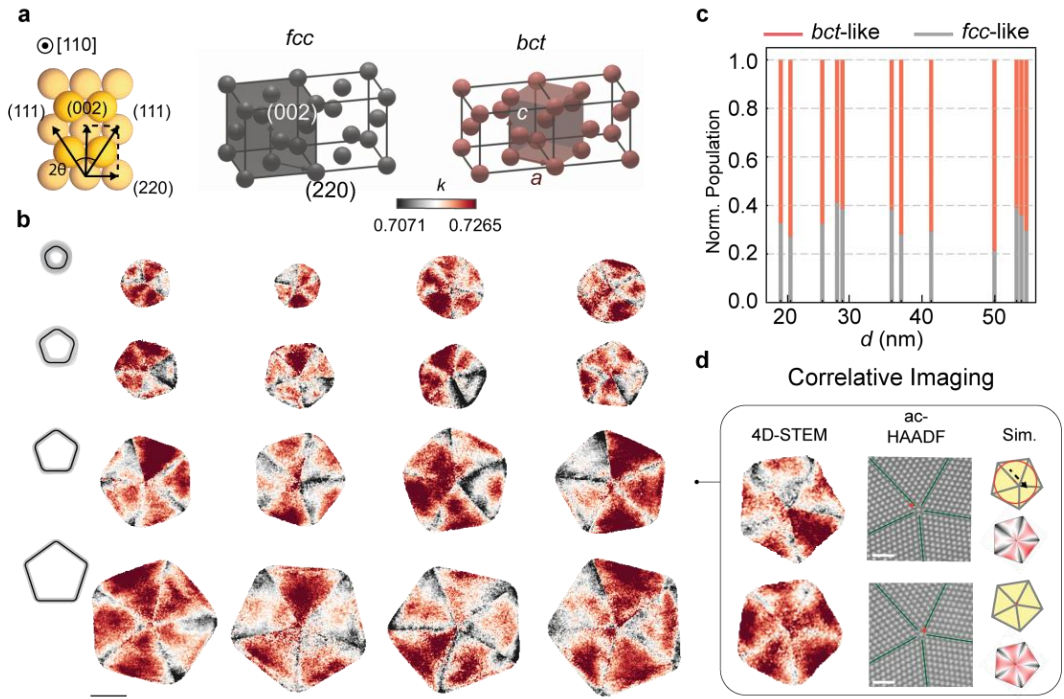

**Fig. 4 Mapping of local lattice distortion of all particles.** (**a**) Left: Au lattice model at (110) orientation with (002), (111), (220) lattice plane directions indicated and θ angle defined. Right: schematics of symmetry transformation when the *d*-spacings of (002) and (220) lattice planes (in arrows) are strained unevenly and how parameter *k* is defined at two extreme cases. (**b**) Lattice distortion maps of all particles, inset on the left showing schematics of variation of particle morphology and tip truncation at each size range. (**c**) Population statistics of *fcc* and *bct* symmetry at different sizes. (**d**) Correlative imaging of two other particles at $d \sim 35$ nm using ac-STEM and 4D-STEM, showing the impact particle center shifting upon the field of lattice distortion. Scale bars: 20 nm (4D-STEM mappings) and 1 nm (HAADF images).

Lastly, we use correlative imaging of 4D-STEM-based lattice symmetry *k* map and HAADF to understand the physical origin of intra-grain heterogeneity. In Figure 4d, two particles at $d \sim 35$ nm show distinctively different mappings of lattice symmetry. The top particle shows one grain with significant lattice distortion and large intra-grain heterogeneity, while the bottom particle shows a symmetrical distribution. From the HAADF images, we estimate the ideal positions of particle center based on the particle shapes by inscribing a circle that approximates the NP shape as in theoretical models[34]. In the two particles, the ideal centers (red dots, Figure 4d) do not fall on the experimentally observed center, suggesting center shifting, but, in the top particle, the shifting is larger than that in the bottom particle. In FEA simulation, when we build models with the center shift and keep all other factors identical, the grain with significantly distorted *k* can be reproduced (top, Figure 4d). This comparison suggests the feature of intra-grain heterogeneity is very sensitive to the ideal position of the disclination axis. Physically, when the center shifts along the direction of a twin boundary, it increases the interplanar twin boundary angle, 2θ, of the opposite grain and thus increases symmetry distortion of the entire grain. When we measure all five interplanar twin boundary angles for all grains, we find 2θ qualitatively match with the severity of lattice symmetry distortion in individual grain for each particle, suggesting a correlation between intra-grain heterogeneity and the shifting of particle centers which is related to their shapes.

**Discussion**

Although inter-grain heterogeneity has been observed as image contrast in TEM as early as 1987 and also frequently in modern reports of Dh NP synthesis, it has remained experimentally unexplored prior to our work[35]. Theoretical studies by Gryaznov *et al.* proposed this "center-shifting" or "disclination core displacement" as one of the seven defect mechanisms of disclination (additional twin boundaries, incoherent surface layers, etc.)[34]. Our work demonstrates a correlation between the center shifting and internal structure of the entire particle and suggests the resulting nanoscale heterogeneity may facilitate particle to grow larger. Regarding the energetic impact of this defect, Ajayan and Marks theoretically showed the energy cost of shifting particle center can be modeled by a parabolic function of the amount of shifting, which has a net gain in strain energy, at the cost of increased total surface energy[36]. Consequently, the shape of each grain may no longer follow the simple modified Wulff shape.[37] Unlike these idealized model with circular shapes and five twin boundaries distributed evenly at particle surfaces, our results show, in realistic colloidal synthesis, particle geometry is much more complex. This includes the convoluted relationship among shape, disclination axis, spatial distribution of twin boundaries. It requires more geometric descriptors to understand the internal structure of geometrically frustrated NPs and the nanoscale heterogeneity.

For the size dependence of lattice straining when adapting atomic geometric frustration, we speculate it is a combination of results from the following factors. First, when particle shape evolves from a rounded shape to faceted pentagonal bipyramidal shape as particle size increases, the strain field distribution changes. This trend may vary by the final Dh particle shape that is dictated by composition, ligands, and surface energy ratio between (111) and (100) ($\gamma_{111}/\gamma_{100}$)[38]. Second, there is significant heterogeneity at nanoscale at intra-grain and inter-grain levels that changes with particle size as discussed above, related possibly due to the center shifting aspects. We observe a crossover of the heterogeneity levels of the $\varepsilon$ strain components at $d \sim 35$ nm, which coincides with the quantitative analysis of geometric gap closing using $\gamma$ and $R$ maps. Following the particle's crystallography, we visualize the transition matches with the emergence of regions of *fcc*-like local lattice symmetry in a single grain, which, upon increasing particle size, distributes to areas near twin boundaries and homogenizes towards bulk behaviors.

Perhaps the biggest surprise is how mechanically soft the Au lattice is, compared to other nanocrystalline metals such as Ag, Pd, and Cu, that show extended defects to sustain the geometric frustration[39–41]. In other words, our workflow of mapping and understanding pseudosymmetry can extend to the multi-twinned NPs of different compositions to obtain more comprehensive understanding at the nanoscale. In the Au Dh NPs, although local defects such as vacancies may exsit[42], they are unlikely contribute significantly to the long-range features of strain fields as extended defects (e.g., nanotwins, dislocation)[43]. Specifically, using the five-fold twins in Ag NPs as a comparison which are more defect-prone, Au exhibits a distinctive set of bulk mechanical properties: doubled intrinsic stacking-fault energy (32, Au vs. 16, Ag, units in mJ/m$^2$), doubled twin-boundary energy (15 vs. 8), yet half the grain boundary energy (364 vs. 790)[44]. These differences suggest mechanically five-fold twinned structure of Au is more stable due to the low energy cost to generate twin boundary to accommodate the geometric frustration while preserving the Dh shape. These distinctions in mechanical properties help to explain previously reported studies on Ag Dh NP, where particle shapes can change when twin boundaries annihilate with particle surfaces and create new facets[39].

More broadly, the mechanisms by which soft lattice accommodates symmetry breaking through shearing and twinning can be generalized to the phase transformation in metallurgy and ceramics.

For example. martensitic transformation is fundamental process in metallurgy that undergoes *fcc* to *bcc/bct* phase transformation upon quenching the materials[45]. Though Bain distortion describe the homogeneous framework of pathways, it alone is insufficient to account for the observed crystallographic orientation relationships, which require consideration of additional lattice-invariant deformation pathways such as twinning or slip[46]. Our work demonstrates diffraction-based 4D-STEM can decouple lattice shearing and phase transformation in a twinned system under special twin geometry and surface constrains of the five-fold Dh NP. We envision the method can be widely adaptable to metals or ceramics to not only study the fundamental pathways by local crystallography with nm spatial resolution but also provide the opportunity to study interaction with local microstructure for mechanical and functional properties.

## Methods

### Chemicals and Materials.

Tetrachloroauric acid trihydrate ($HAuCl_4 \cdot 3H_2O$, $\geq$ 99.9%), hexadecyltrimethylammonium chloride (CTAC) solution at 25 wt.% in $H_2O$, benzyldimethylhexadecylammonium chloride (BDAC), citric acid ($\geq$ 99.5%), sodium borohydride ($NaBH_4$, $\geq$ 99.0%), L-ascorbic acid (AA, $\geq$ 99.9%) were all purchased from Sigma Aldrich and used without further purification. Deionized (DI) water with a resistivity of 18.2 MΩ·cm (at room temperature) was used throughout the experiments.

### Synthesis of gold seeds.

Gold seeds were prepared in a scintillation vial (20 mL) by mixing $HAuCl_4$ (0.25mM) with citric acid (5 mM) in an aqueous CTAC solution (10 mL, 50 mM) under vigorous stirring (1,200 rpm) at room temperature. Freshly prepared $NaBH_4$ (0.25 mL, 25 mM, with iced water) was then added, followed by loosely capping the vial to avoid quick oxidation of $NaBH_4$. The solution immediately turned from yellow to brown, indicating the formation of Au clusters. After 2 minutes, the vial was closely capped and heated in an oil bath at 90 °C for 90 minutes under gentle stirring (400 rpm). The color gradually changed from brown to red, suggesting the formation of Au seeds. Afterwards, the seed solution was taken out from oil bath, quench to room temperature, and stored in ambient condition.

### Synthesis of gold decahedral nanoparticles.

A desired volume of gold seed solution was added to a growth solution containing $HAuCl_4$ (0.1 mL, 50 mM), BDAC (10 mL, 100 mM), AA (0.075mL, 100 mM) under vigorous stirring at 30 °C. Afterwards, the mixture was left undisturbed at 30 °C for 30 minutes, and the products were collected by centrifugation (8,000 rpm, 10 min). The sizes of decahedral nanoparticles were controlled by the volume of gold seeds added: 500 μL for 20–30 nm NPs, 200 μL for 30–40 nm NPs, and 100 μL for 40–50 nm NPs.

### 4D-STEM measurement

4D-STEM and micro-probe (μp) STEM imaging was performed on a ThermoFisher Talos S/TEM with FEG-X gun operating at 200 kV in μp mode. With 20 μm condenser (C2) aperture, gun lens 4–5, spot size 7, the probe condition can repeatedly reproduce with convergence angle of 0.6 mrad, beam size of 2.0 nm at full-width-at-maximum (FWHM), and beam current of 7.4 pA (or $1.5 \times 10^5$ $e^- Å^{-2} s^{-1}$). 4D-STEM datasets were collected on an Electron Microscope Pixelated Array Detector (EMPAD) at nominal camera length of 520 mm and nominal magnification of 1.05Mx with the step size of 0.376 nm and exposure time of 1 ms. The HAADF images were collected by the same probe and imaging conditions using a HAADF detector with dwell time of 2–4 µs. Dh nanoparticles were carefully tilt to the common [110] zone axis that all five grains share before raster scan of electron beam started. All particles were maintained at the eucentric plane for a consistent sample-to-detector distance across particle for quantitative comparison. The size of each 4D-STEM dataset is 256 × 256 pixels as such the data collections took ~1 min.

**Corresponding Author**
*Qian Chen, Email: qchen20@illinois.edu

**Author contributions**
O.L. and Q.C. conceived the project. Z.L. synthesized the materials. O.L. and C.-Y.H. conducted the experiments. O.L., H.-C.Ni, X.W., C.-Y.H., and L.Y. carried out and streamlined experimental data analysis. Y.J. performed the FEA simulations. Q.C. and J.-M.Z. supervised the work. O.L. and Q.C. wrote the manuscript. All authors discussed the results and contributed to the final manuscript.

**Acknowledgements**
This research is based upon work supported by the U.S. Department of Energy, Office of Science, Office of Basic Energy Sciences, under Award Number DE-SC0024064. Experiments were carried out in part in the Materials Research Laboratory Central Research Facilities, University of Illinois.

**Competing interests**
The authors declare no competing interests.